\documentclass[a4paper,11pt]{article}
\usepackage{jcappub, hyperref, cleveref}
\usepackage{amsmath}
\usepackage{graphicx}   % Figure
\graphicspath{{Fig/}}

\usepackage{xcolor}
\usepackage[normalem]{ulem}
\newcommand{\ii}{{\rm i}}

\begin{document}

\title{Circularly Polarized Gravitational Wave Background Search with a Network of Space-borne Triangular Detectors}

%Parity-violating 
%\author{}\email{}
\author{Ju Chen$^{1,5}$} %\email{chenju@ucas.ac.cn}
\author{Chang Liu$^{2,*}$} %\email{liuchang@yzu.edu.cn}
\author{Yun-Long Zhang$^{3,4,*}$ 
\note[*]{Corresponding authors}
}

\emailAdd{chenju@ucas.ac.cn}
\emailAdd{liuchang@yzu.edu.cn}
\emailAdd{zhangyunlong@nao.cas.cn}

\affiliation{$^{1}$International Center for Theoretical Physics Asia-Pacific (ICTP-AP), University of Chinese Academy of Sciences, Beijing 100190, China}

\affiliation{$^{2}$Center for Gravitation and Cosmology, College of Physical Science and Technology, Yangzhou University, Yangzhou, 225009, China}

\affiliation{$^{3}$National Astronomical Observatories, Chinese Academy of Sciences, Beijing, 100101, China}

\affiliation{$^{4}$School of Fundamental Physics and Mathematical Sciences,  Hangzhou   Institute for Advanced Study, UCAS, Hangzhou 310024, China.}

\affiliation{$^{5}$Taiji Laboratory for Gravitational Wave Universe (Beijing/Hangzhou), University of Chinese Academy of Sciences, Beijing 100049, China}

\date{\today}

\allowdisplaybreaks

%%%%%%%%%%%%%%%%%%%%%%%%%%%%%%%%%%%%%%%%%%%%%%%%%%%%%%%%%%%%%%%%%%%%%%%%%%%%
\abstract{Circularly polarized gravitational wave backgrounds are predicted in many well-motivated models of inflation and phase transitions involving spontaneous parity violation. In this work, we investigate the detection of such parity-violating signals with the network of two space-borne triangular detectors. We derive the general analytical formula for the overlap reduction functions of networks composed of two triangular detectors, by exploiting the system's symmetrical properties under the long-wave approximation. Based on these results, we further assess the detectability of a circularly polarized background with alternative LISA-Taiji network configurations. We find that the sensitivity to the parity-violating component can be significantly enhanced at low frequencies by adjusting the orientation of Taiji's constellation plane. This sensitivity gain is approximately an order of magnitude around the millihertz band, making the peak sensitivity comparable to that of the total intensity. This provides a promising opportunity to constrain various parity-violating theories in the millihertz band with upcoming space-borne detectors.}

\maketitle
\tableofcontents
%%%%%%%%%%%%%%%%%%%%%%%%%%%%%%%%%%%%%%%%%%%%%%%%%%%%%%%%%%%%%%%%%%%%%%%%%%%%

\section{Introduction}
The stochastic gravitational-wave background (SGWB) is one of the most intriguing topics in gravitational-wave research. Due to the inherently weak nature of gravitational interactions, detecting such background is highly challenging. However, it also presents a unique opportunity to directly probe the early universe beyond the observational limitations of the cosmic microwave background (CMB) \citep{Hogan2006, Allen1996, Caprini2018}. This would provide invaluable insights into the history of the early universe that is otherwise unattainable through electromagnetic observations.
% last-scattering surface of the cosmic microwave background.
Various mechanisms for the generation of cosmological gravitational waves (GWs) have been studied, including quantum fluctuations during inflation \citep{Grishchuk1975, Starobinsky1979}, bubble collisions during first-order phase transitions \citep{Witten1984, Kamionkowski1994}, cusps and kinks of cosmic strings \citep{Damour2001, Siemens2007}, and the turbulence of primordial magnetic plasma \citep{Kosowsky2002, Dolgov2002}. Additionally, many astrophysical processes, such as close compact binaries \citep{Schneider2001, Farmer2003} or rapidly rotating neutron stars \citep{Ferrari1999, Regimbau2001}, are also expected to contribute to this background. A comprehensive review of the various sources of the SGWB can be found in \citep{Christensen2019}.

The SGWB is commonly assumed to be unpolarized due to the stochastic and uncorrelated nature of the generation process. However, parity violation in gravity could alter both the generation and propagation of gravitational waves, resulting in a circularly polarized SGWB. In fact, a significant degree of polarization can be generated in well-motivated models of inflation with spontaneous parity violation, such as leptogenesis \citep{Alexander2006} or Chern-Simons coupling \citep{Satoh2008,Adshead2012}. It has also been shown that, in the presence of helical primordial magnetic fields \citep{Cornwall1997,Vachaspati2001}, turbulence in the magnetic plasma during first-order phase transition could also give rise to a circular polarized SGWB \citep{Kahniashvili2005,RoperPol2020}. Additionally, parity violations may induce birefringence during the propagation of gravitational waves \citep{Callister2023a,Li2024}, leading to a net circular polarization in the background.

Due to the potential cosmological origin, parity-violating background signals could manifest in various frequency bands. The (non-)detection of such a background would offer a novel way to explore the nature of gravity and place constraints on various parity-violating theories. Many efforts have been made using different gravitational wave detectors. For ground-based detectors, current sensitivity remains insufficient to place significant restrictions on circular polarization \citep{Martinovic2021, Jiang2023a}, though measurements are anticipated in the coming years \citep{Omiya2023a}. For space-borne detectors, several preliminary predictions for the LISA-Taiji network have been presented in recent studies \citep{Seto2020, Orlando2021}, suggesting that detection limits could reach $10^{-12}$ in terms of fractional energy density with ten years of observation \citep{Seto2020}. Pulsar Timing Arrays (PTAs), while insensitive to circular polarization of an isotropic background, have been shown to potentially detect such parity-violating signal in an anisotropic background \citep{Kato2016, Belgacem2020a}.

For an isotropic SGWB, circular polarization is undetectable with planar detector(s) \citep{Seto2006,Seto2007}. The detection of such signals requires correlating gravitational wave detectors that are not co-planar, which means a network of detectors is necessary. In this paper, we study the detection of a parity-violating isotropic SGWB with a network of two space-borne triangular detectors under the long-wave approximation. We derive the general analytical formula for the overlap reduction functions of a network with two triangular detectors, utilizing the system's symmetry following \cite{Seto2020}. As a demonstration, we further assess the detectability of parity-violating SGWB using the alternative LISA-Taiji networks proposed in \citep{Wang2021c}.

In the following section~\ref{sec:general}, we detail the detection of parity-violating SGWB and give the analytical formulas of effective overlap reduction functions for the network of space-borne detectors. In section~\ref{sec:network}, we focus on the (alternative) LISA-Taiji network and discuss on more general configurations of the network. The conclusions are presented in section \ref{sec:conclusion}.

\section{Circularly Polarized Gravitational Wave Background}
\label{sec:general}

The SGWB is formed as a superposition of plane waves from all possible directions and at various frequencies \citep{Maggiore2007}:
\begin{equation}
  h_{ab}(t, \vec{r}) = \sum_{P=+ , \times}\int_{-\infty}^{\infty} df \int d^2\hat{k} ~ \tilde{h}_P(f,\hat{k}) e_{ab}^P(\hat{k}) e^{\ii 2\pi f(t - \hat{k}\cdot\vec{r}/c)},
\end{equation}
where $e^P_{ab}(\hat{k})$ represents the basis tensor for a specific polarization $P$ and propagation direction $\hat{k}$.
$\tilde{h}_P(f,\hat{k})$ is the Fourier transform of the corresponding plane waves. Here, we consider the two polarization modes in general relativity, $P =+, \times$.
Assuming the SGWB is stationary and isotropic, with possible nonzero net polarization, the ensemble average of the Fourier amplitudes can be expressed as \citep{Seto2006}:
\begin{equation}
\label{eq:stocks}
\begin{aligned}
\begin{pmatrix}
   \langle \tilde{h}_+(f, \hat{k}) \tilde{h}_+^*(f', \hat{k}')\rangle & \langle \tilde{h}_+(f, \hat{k}) \tilde{h}_\times^*(f', \hat{k}')\rangle \\
   \langle \tilde{h}_+^*(f, \hat{k}) \tilde{h}_\times(f', \hat{k}')\rangle & \langle \tilde{h}_\times(f, \hat{k}) \tilde{h}_\times^*(f', \hat{k}')\rangle
\end{pmatrix} = \frac{1}{2}\delta(f-f') \frac{\delta^2(\hat{k}-\hat{k}')}{4\pi}
\begin{pmatrix}
   ~I(f)  & -iV(f) \\
   iV(f) & ~~I(f)
\end{pmatrix},
\end{aligned}
\end{equation}
where $I$, $V$ are the Stokes parameters, with $I$ related to the overall intensity of the background and $V$ characterizes the asymmetry of the intensity of right and left-hand circular polarization, i.e. the parity-violating components. 
The other two Stokes parameters, \( Q \) and \( U \), which describe linear polarization, are spin-4 quantities and do not contribute to the monopole modes, i.e., the isotropic components of the stochastic background. More detailed discussions can be found in \citep{Seto2006, Seto2007a, Liu:2022umx}. In this work, we have omitted these two components.

It is more convenient to introduce the basis tensors for circular polarization modes:
\begin{equation}
\label{eq:circular_basis}
e^R_{ab} = \frac{e^+_{ab} + i e^\times_{ab}}{\sqrt{2}}, ~~~ e^L_{ab} = \frac{e^+_{ab} - i e^\times_{ab}}{\sqrt{2}},
\end{equation}
and the corresponding strains:
\begin{equation}
\label{eq:circular_strain}
h_R = \frac{h_+ - i h_\times}{\sqrt{2}}, ~~~~~ h_L = \frac{h_+ + i h_\times}{\sqrt{2}}.
\end{equation}
Using this notation, the ensemble average in Eq.~\eqref{eq:stocks} becomes
\begin{align}
&\langle\tilde{h}_\lambda(f, \hat{k})\tilde{h}^*_{\lambda'}(f', \hat{k}')\rangle = \frac{1}{2}\delta(f-f') \frac{\delta^2(\hat{k}-\hat{k}')}{4\pi}  \delta_{\lambda \lambda'} S_\lambda(f),
\end{align}
where $\lambda=R, L$, and $S_R(f)$ and $S_L(f)$ are the spectral densities of right and left-hand circular polarization, respectively. These spectral densities are related to the Stokes parameter as follows:
\begin{equation}
  \begin{aligned}
     I(f) = \frac{1}{2}[S_R(f) + S_L(f)], ~~~  V(f) = \frac{1}{2}[S_R(f) - S_L(f)].
  \end{aligned}      
\end{equation}
Notice that comparing with Eq.~(8.38) in \cite{Romano2017}, there is an additional factor 1/2 due to different definitions of $I(f)$.
From Eq. (2.15) in \cite{Romano2017},
$I(f)\equiv S_h(f)$ is defined as the summation over both polarizations. 
In our case, the definition of $I(f)$ corresponds to the mean of both polarizations. We have chosen this convention in \eqref{eq:stocks}, which is consistent with those in \cite{Seto2006, Maggiore2007}, such that $\Omega_{\text GW} = \frac{4\pi^2 f^3 }{3H_0^2}I(f)$.

\subsection{Detection of circularly polarized background}

The space-borne interferometer like LISA \citep{Amaro-Seoane2017} operate as synthetic interferometers. Various combinations of the readouts from each inter-spacecraft laser link, known as time-delay interferometry (TDI) channels, are synthesized in post-processing to effectively suppress laser noise.  
Commonly, the noise-orthogonal channels, $A$ and $E$, are constructed under the assumption of equal arm lengths, identical and uncorrelated noise in each laser link \citep{Prince2002, Adams2010}.  
In this work, we focus on the low-frequency regime, 
the response of a space-borne detector to the SGWB signal in the frequency domain is expressed as:
\begin{eqnarray}
\tilde{s}(f) &=& \frac{\sqrt{3}}{2} \sin\frac{2\pi f L}{c} D^{ab} \tilde{h}_{ab}(f, \vec{r}),
\end{eqnarray}
where $L$ represents the arm lengths of the detector, and $D^{ab}$ is the detector tensor for the $A$ or $E$ channel (see Appendix~\ref{sec:AET} for further details). $\tilde{h}_{ab}(f, \vec{r})$ is the Fourier transform of the SWGB signal $h_{ab}(t, \vec{r})$ at the detector location  $\vec{r}$. With the circular polarization basis tensors introduced earlier:
\begin{eqnarray}
  \tilde{h}_{ab}(f, \vec{r}) &=& \sum_{\lambda=R , L} \int d^2\hat{k} ~ \tilde{h}_\lambda(f,\hat{k}) e_{ab}^\lambda(\hat{k}) e^{-\ii 2\pi f\hat{k}\cdot\vec{r}/c},
\end{eqnarray}
where $e_{ab}^\lambda(\hat{k})$ is the circular polarization basis tensor as defined in Eq.~\eqref{eq:circular_basis}, and $ \tilde{h}_\lambda(f,\hat{k})$ represents the corresponding Fourier amplitudes provided in Eq.~\eqref{eq:circular_strain}.

The detection of parity-violating background requires the correlation of detectors that are non-coplanar. For two detectors (or channels), labeled $i$ and $j$, the correlation of their signals is given by \citep{Seto2006}:
  \begin{equation}
  \label{eq:correlation}
  \langle \tilde{s}_i(f) \tilde{s}^*_j(f')\rangle = \frac{1}{2}\delta(f-f') \left[\Gamma^I_{ij}(f)I(f) + \Gamma^V_{ij}(f)V(f) \right],
    \end{equation}
where $\Gamma^I_{ij}$ and $\Gamma^V_{ij}$ are the overlap reduction functions (ORFs) of the intensity $I$ and parity-violating $V$ components of the isotropic SGWB, respectively. These ORFs  describe how the responses of the two detectors are correlated, depending on their separation and orientation. Using Eqs.~\eqref{eq:circular_basis}-\eqref{eq:correlation},  the ORFs can be expressed as:
\begin{equation}
  \begin{aligned}
    \label{eq:overlap}
    \Gamma^I_{ij}(f) &\equiv \frac{3}{4} \sin\frac{2\pi f L_i}{c} \sin \frac{2\pi f L_j}{c} D_i^{~ab} D_j^{~cd} \int \frac{d^2\hat{k}}{4\pi} ~ \left[e^+_{ab}(\hat{k}) e^+_{cd}(\hat{k}) + e^\times_{ab}(\hat{k}) e^\times_{cd}(\hat{k})\right]  e^{-\ii 2\pi f\hat{k}\cdot \Delta\vec{r}/c},\\
    \Gamma^V_{ij}(f) &\equiv  -\ii ~\frac{3}{4} \sin\frac{2\pi f L_i}{c} \sin \frac{2\pi f L_j}{c}  D_i^{~ab} D_j^{~cd} \int \frac{d^2\hat{k}}{4\pi} \left[e^+_{ab}(\hat{k}) e^\times_{cd}(\hat{k}) - e^\times_{ab}(\hat{k}) e^+_{cd}(\hat{k})\right]  e^{-\ii 2\pi f\hat{k}\cdot \Delta\vec{r}/c}, 
  \end{aligned}
\end{equation}
where $\Delta \vec{r} \equiv \vec{r}_i-\vec{r}_j$ is the separation vector between the two detectors.

It is more convenient to introduce the normalized ORFs, $\gamma^I_{ij}$ and $\gamma^V_{ij}$, which removes the influence of the detector response in the long-wave limit and accounts solely for the effects of the separation and misalignment of the detectors.
For $\Gamma^I_{ij}(f)$, the normalization is chosen such that $\gamma^I_{ij}(0) = 1$ for a co-located and co-aligned detector pair \citep{Cornish2001}, which corresponds to $\frac{3}{10} \sin\frac{2\pi f L_i}{c} \sin\frac{2\pi f L_j}{c}$ in the long-wave limit. For $\Gamma^V_{ij}(f)$,  the same normalization factor is applied as for the $I$ component: 
\begin{equation}
\begin{aligned}
  \label{eq:orf}
  \gamma^I_{ij}(f) &= \frac{10}{3}\frac{\Gamma^I_{ij}(f) }{\sin(2\pi f L_i/c) \sin(2\pi f L_j/c)}, ~~~\\ \gamma^V_{ij}(f) &= \frac{10}{3}\frac{\Gamma^V_{ij}(f) }{\sin(2\pi f L_i/c) \sin(2\pi f L_j/c)}.
    \end{aligned}
\end{equation}
The calculation of $\gamma^I_{ij}$ can be carried out analytically, as shown in \citep{Flanagan1993,Allen1999}. A similar approach can be used to derive the analytic form of $\gamma^V_{ij}$. More details on these calculations are present in Appendix~\ref{sec:ORFtensor}.

To unambiguously identify the parity-violating components, it is necessary to separate the $I$ and $V$ components from the cross-correlation of the detected signals, as expressed in Eq.~\eqref{eq:correlation}. 
Considering that each channel pair yields a linear combination of $I$ and $V$, the two components can be solved from the linear system of multiple pairs. The signal-to-noise ratio (SNR) can be determined using Fisher's method. A detailed analysis can be found in Section 8.2.3 of \cite{Romano2017} (see also \citep{Seto2007, Seto2008}).
With two space-borne triangular detectors, there are four independent channel pairs, which we denote as $\kappa \in \{A_1{\text -}A_2, A_1{\text -}E_2, E_1{\text -}A_2, E_1{\text -}E_2\}$. For each detector the noise power spectral density (PSD) of $A$ and $E$ channels is identical, thus all the four channel pairs exhibit the same noise characteristics. Under this condition, the expression for the optimal signal-to-noise ratio (SNR) can be significantly simplified, as demonstrated in \citep{Seto2008}:
\begin{equation}
  \begin{aligned}
    \label{eq:SNR}
    SNR_I^2 &=\left(\frac{3}{10}\right)^2 T_{\rm obs} \int df \left[\sum_\kappa (\gamma^I_\kappa)^2 - \frac{\big(\sum_\kappa \gamma^I_\kappa\gamma^V_\kappa\big)^2}{\sum_\kappa (\gamma^V_\kappa)^2}\right] I^2/N^2,\\
    SNR_V^2 &=\left(\frac{3}{10}\right)^2 T_{\rm obs} \int df \left[\sum_\kappa (\gamma^V_\kappa)^2 - \frac{\big(\sum_\kappa \gamma^I_\kappa\gamma^V_\kappa\big)^2}{\sum_\kappa (\gamma^I_\kappa)^2}\right] V^2/N^2 .     
  \end{aligned}
\end{equation}
Here $N \equiv \sqrt{N_1(f) N_2(f)/[\sin(2\pi f L_1/c) \sin(2\pi f L_2/c)]}$, $N_1(f)$ and $N_2(f)$ represent the noise PSD of the $A$ or $E$ channels for each detector, respectively. 

\subsection{Effective overlap reduction functions}
\label{sec:ORF}
It is convenient to introduce the effective ORFs:
 \begin{equation}
  \begin{aligned}
    \label{eq:orf_eff}
    \gamma_{\rm eff}^I = \sqrt{\sum_\kappa (\gamma^I_\kappa)^2 - \frac{\big(\sum_\kappa \gamma^I_\kappa\gamma^V_\kappa\big)^2}{\sum_\kappa (\gamma^V_\kappa)^2}}, ~~~~
    \gamma_{\rm eff}^V = \sqrt{\sum_\kappa (\gamma^V_\kappa)^2 - \frac{\big(\sum_\kappa \gamma^I_\kappa\gamma^V_\kappa\big)^2}{\sum_\kappa (\gamma^I_\kappa)^2}}\,.    
  \end{aligned}
\end{equation}
The optimal SNR is then fully determined by the effective ORFs $\gamma_{\rm eff}^I$ and $\gamma_{\rm eff}^V$, along with the SWGB spectral density $I(f)$ and $V(f)$.

The effective ORFs depends on the summation of ORFs over all channel pairs: $\sum_\kappa (\gamma^I_\kappa)^2$, $\sum_\kappa (\gamma^V_\kappa)^2$ and $\sum_\kappa \gamma^I_\kappa\gamma^V_\kappa$. Here we first derive the individual $\gamma^I_\kappa$ and $\gamma^V_\kappa$. Following \citep{Flanagan1993,Allen1999}, with the auxiliary tensors $\hat{\Gamma}_{abcd}^I$ and $\hat{\Gamma}_{abcd}^V$ introduced in Appendix~\ref{sec:ORFtensor}, the normalized ORFs can be formed as:
\begin{equation}
  \label{eq:orf_decomposed}
  \gamma^I_{ij} = D^{ab}_i D^{cd}_j \hat{\Gamma}^I_{abcd}, ~~~ \gamma^V_{ij} = D^{ab}_i D^{cd}_j \hat{\Gamma}^V_{abcd}.
\end{equation}
The tensors $\hat{\Gamma}_{abcd}^I$ and $\hat{\Gamma}_{abcd}^V$ are expressed in terms of Kronecker's delta $\delta_{ab}$, the unit separation vector $s^a$ and the anti-symmetric tensor $w_{ab}\equiv \epsilon_{abc} s^c$ as follows:
\begin{equation}
 \begin{aligned}
  \hat{\Gamma}^I_{abcd} &= b^I_0 \delta_{ac}\delta_{bd} + b^I_1 \delta_{ac} s_{b} s_{d} + b^I_2 s_{a} s_{b} s_{c} s_{d}, \\ \hat{\Gamma}^V_{abcd} &= b^V_0 w_{ac}\delta_{bd} + b^V_1 w_{ac} s_{b} s_{d}, 
   \end{aligned}
\end{equation}
where the coefficients $b^I_0, b^I_1, b^I_2$ and $b^V_0, b^V_1$ are defined by the spherical Bessel functions:
\begin{equation}
  \label{eq:Bessel}
  \begin{aligned}
    b_0^I(\iota) &= 2j_0 - \frac{20}{7}j_2 + \frac{1}{7}j_4, ~~~~~ b_1^I(\iota) = \frac{60}{7}j_2 - \frac{10}{7}j_4, ~~~~~ b_2^I(\iota) = \frac{5}{2}j_4,\\
    b_0^V(\iota) &= 4j_1 - j_3, \qquad \qquad ~~~~~ b_1^V(\iota) = 5j_3.
  \end{aligned}
\end{equation}
Here $j_0, j_1, j_2, j_3, j_4$ are spherical Bessel functions with argument $\iota = 2\pi f \Delta r/c$. Further details are provided in Appendix~\ref{sec:ORFtensor}. The normalized ORFs $\gamma^I_{ij}$ and $\gamma^V_{ij}$ for a given pair of detectors (or channels) are then obtained by contracting the detector tensors with the auxiliary tensors, following Eq.~\eqref{eq:orf_decomposed}. As scalar functions, the ORFs can be calculated for each channel pair with any chosen coordinate system without loss of generality. In general, $\gamma_{ij}$ depends on the distance and relative orientation of the different detectors (or channels), which are characterized by five angles, as outlined in \citep{Flanagan1993}. The resulting ORFs typically vary due to the ``cartwheel motion'' of space-borne detectors. 

The summations of ORFs $\sum_\kappa (\gamma^I_\kappa)^2$, however, are invariant under the rotation of detectors within their respective constellation planes, as highlighted in \citep{Seto2020}. The same conclusion applies to $\sum_\kappa (\gamma^V_\kappa)^2$ and $\sum_\kappa \gamma^I_\kappa\gamma^V_\kappa$, and these summations can be determined by leveraging the symmetry of the system. To understand this, we first note that each individual ORFs can be decomposed into two parts following Eqs.~\eqref{eq:orf_decomposed}-\eqref{eq:Bessel}: an oscillatory part, characterized by the coefficients provided in Eq.~\eqref{eq:Bessel}, which encapsulate frequency-dependent behavior, and a geometric part determined by the detector tensors $D^{ab}_i$ of each detector (or channel) and unit separation vector $s^a$ between the two detectors:
\begin{equation}
  \label{eq:orf_factored}
  \gamma_{ij}^I = \sum_{m=0}^2 b_m^I ~\chi_{ij}^{Im}, ~~~~~~~
  \gamma_{ij}^V = \sum_{m=0}^1 b_m^V ~\chi_{ij}^{Vm},
\end{equation}
where
\begin{equation}
  \begin{aligned}
    \chi_{ij}^{I0} &= D^{ab}_i D^{cd}_j\delta_{ac}\delta_{bd}, ~~~~ \chi_{ij}^{I1} = D^{ab}_i D^{cd}_j\delta_{ac}s_b s_d, ~~~~ \chi_{ij}^{I2} = D^{ab}_i D^{cd}_j s_a s_b s_c s_d, \\
    \chi_{ij}^{V0} &= D^{ab}_i D^{cd}_j w_{ac}\delta_{bd}, ~~~~ \chi_{ij}^{V1} = D^{ab}_i D^{cd}_j w_{ac}s_b s_d.
  \end{aligned}
\end{equation}
From this, the summations of ORFs can also be factored into oscillatory and geometric parts. Formally we have:
\begin{equation}
\label{eq:orf_summations}
  \begin{aligned}  
  % Y^I &\equiv
  \sum_\kappa (\gamma^I_\kappa)^2 = \sum_{m=0}^{2}\sum_{n=0}^{2} b^I_m(\iota) ~ b^I_n(\iota)~ X^I_{mn},\\
  % Y^V &\equiv
  \sum_\kappa (\gamma^V_\kappa)^2 = \sum_{m=0}^{1}\sum_{n=0}^{1} b^V_m(\iota)~ b^V_m(\iota)~ X^V_{mn},\\
  % Y^{IV} &\equiv
  \sum_\kappa \gamma^I_\kappa \gamma^V_\kappa = \sum_{m=0}^{2}\sum_{n=0}^{1} b^I_m(\iota)~ b^V_n(\iota)~ X^{IV}_{mn},
  \end{aligned}
\end{equation}
where
\begin{equation}\label{eq:Xmn}
  X^I_{mn} = \sum_\kappa \chi_\kappa^{Im} \chi_\kappa^{In}, ~~~ X^V_{mn} = \sum_\kappa \chi_\kappa^{Vm} \chi_\kappa^{Vn}, ~~~ X^{IV}_{mn} =\sum_\kappa \chi_\kappa^{Im} \chi_\kappa^{Vn}, 
\end{equation}
are scalar quantities that depend only on the geometry of the two detectors. Specifically, they are determined by the detector tensors $D^{ab}_A, D^{ab}_E, D^{ab}_{A_1}, D^{ab}_{E_1}$ and the unit separation vector $s^a$. 
%$$\red{\sum_{ij} (D_i^{ab}D_j^{cd})(D_i^{ef}D_j^{gh})}$$

\begin{figure}[ht]
  \centering
  \includegraphics[width=12cm]{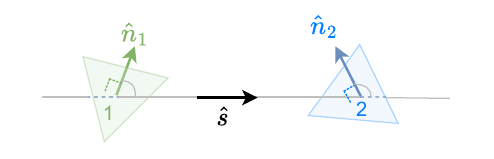} 
  \caption{Schematic diagram of the two triangle detectors' configuration.}
  \label{fig:geometry}
\end{figure}

In fact, all these geometric factors $X_{mn}$ are also invariant under the rotation of detectors within their respective constellation plane, as demonstrated in \citep{Seto2020}. Consequently, all geometric factors $X_{mn}$ depend solely on the unit normal vectors of each detector's constellation plane $\hat{n}_1$ and $\hat{n}_2$, as well as unit separation vector $\hat{s}$ between the two detectors, as illustrated in Fig.~\ref{fig:geometry}, resulting in a total of six degrees of freedom. Additionally, due to the isotropy of the SGWB, all $X_{mn}$ remain unchanged under the overall rotation of the network, effectively eliminating three degrees of freedom. As a result, the geometric factors $X_{mn}$ exhibit only three degrees of freedom, which can be described using the three angles between the unit vectors:
\begin{equation}
  a \equiv \arccos(\hat{n}_1\cdot\hat{s}), ~~~ b \equiv \arccos(\hat{n}_2\cdot\hat{s}), ~~~ c \equiv \arccos(\hat{n}_1\cdot\hat{n}_2).
\end{equation}

Finally, the geometric factors in terms of the three angles can be obtained once a coordinate system is chosen, and the results are present as follows.
The components of parity even geometric part $X^{I}_{mn}$ are
\begin{equation}
  \begin{aligned}
    X^I_{00} &= \frac{1}{16}\left(1+6\cos^2 c + \cos^4 c\right),\\
    X^I_{01} &= X^I_{10} = \frac{1}{16}\left[1-\cos^2 a -\cos^2 b -\cos a \cos b \cos c + \cos a \cos b \cos^3 c\right.\\
& \left.\qquad\qquad\qquad
-(\cos^2 a+\cos^2 b - 3)\cos^2 c\right],\\
    X^I_{02} &= X^I_{20} = \frac{1}{16}\left[ (1+\cos^2 a)(1+\cos^2 b)\cos^2 c - 4\cos a \cos b \cos c(\cos^2 a + \cos^2 b) \right.\\
    & \left.\qquad\qquad \qquad +  (2\cos^2 a + \cos^2 b -1)(\cos^2 a + 2\cos^2 b -1)\right],\\
    X^I_{11} &= \frac{1}{16}\left(1+\cos^2 c\right)\sin^2 a \sin^2 b,\\
    X^I_{12} &= X^I_{21} = \frac{1}{16}\left(1-\cos^2 a - \cos^2 b + \cos a \cos b \cos c\right)\sin^2 a \sin^2 b,\\
    X^I_{22} &= \frac{1}{16} \sin^4 a \sin^4 b\,.
  \end{aligned}
\end{equation}

The components of parity odd geometric part $X^{V}_{mn}$ are
\begin{equation}
\begin{aligned}
X^V_{00} &= \frac{1}{16} \left[2 \cos a ~ \cos b ~ \cos c (3 +\cos^2 c)+ (1 + \cos^2 c)\sin^2c\right],\\
% X^V_{01} &= X^V_{10} = \frac{1}{16} \Big[ \cos a \cos b \cos c ~ ( 5 - 2\cos^2 a - 2\cos^2 b - \cos^2 c)    \\
% & \qquad\qquad\qquad  -\frac{\sin^2 c}{2} \big(2\cos 2a +2 \cos 2b + \cos 2 a \cos 2 b + 1\big) \Big],\\
X^V_{01} &= X^V_{10} = \frac{1}{16} \Big[ \cos a \cos b \cos c ~ {( 2\sin^2 a + 2\sin^2 b + \sin^2 c)}    \\
& \qquad\qquad\qquad  -\frac{\sin^2 c}{2} \big(2\cos 2a +2 \cos 2b + \cos 2 a \cos 2 b + 1\big) \Big],\\
    X^V_{11} &= \frac{1}{16} \sin^2 a \sin^2 b \left(2 \cos a \cos b \cos c+\sin^2c\right)\,.
\end{aligned}
\end{equation}
And, the components of the cross term geometric part $X^{IV}_{mn}$ are 
\begin{equation}
  \begin{aligned}
  X^{IV}_{00} &= -\frac{C}{16} \cos c \left(3 + \cos^2 c\right),\\
%\sqrt{\frac{1}{2}(4\cos a \cos b \cos c - \cos 2a - \cos 2b - \cos 2c - 1)}
% X^{IV}_{01} &= \frac{C}{16} \left[(\cos 2a + \cos 2b) \cos c - \cos a \cos b (-3 + \cos^2 c)\right],\\
X^{IV}_{01} &= \frac{C}{16} \left[(\cos 2a + \cos 2b) \cos c + {\cos a \cos b (3 - \cos^2 c)}\right],
%\sqrt{\frac{1}{2}(4\cos a \cos b \cos c - \cos 2a - \cos 2b - \cos 2c - 1)}
\\  X^{IV}_{10} &= -\frac{C}{16} \left[2\cos c + \cos a \cos b(1+\cos^2 c) \right],
%\sqrt{\frac{1}{2}(4\cos a \cos b \cos c - \cos 2a - \cos 2b - \cos 2c - 1)}
\\  X^{IV}_{11} &= -\frac{C}{16} \sin^2 a \sin^2 b \cos c ,
%\sqrt{\frac{1}{2}(4\cos a \cos b \cos c - \cos 2a - \cos 2b - \cos 2c - 1)}
\\  X^{IV}_{20} &= \frac{C}{16} \left[2\cos a \cos b (\cos^2 a + \cos^2 b) - (1 + \cos^2 a) (1 + \cos^2 b) \cos c\right],
%\sqrt{\frac{1}{2}(4\cos a \cos b \cos c - \cos 2a - \cos 2b - \cos 2c - 1)}
\\  X^{IV}_{21} &=  -\frac{C}{16} \sin^2 a \sin^2 b \cos a \cos b,
%\sqrt{\frac{1}{2}(4\cos a \cos b \cos c - \cos 2a - \cos 2b - \cos 2c - 1)}\\
\end{aligned}
\end{equation}
where 
\begin{equation}
  C \equiv \sqrt{\frac{1}{2}(4\cos a \cos b \cos c - \cos 2a - \cos 2b - \cos 2c - 1)}\,.
\end{equation}

The effective ORFs are then obtained by following Eq.~\eqref{eq:orf_eff}, with the summations of ORFs provided in  Eq.~\eqref{eq:orf_summations}. Despite the variation of the individual ORFs for each channel pair during the orbit, the effective ORFs remain the same,  depending only on the separation of the two detectors and relative orientation of the constellation planes, i.e., the overall network configuration. 

\section{Comparison of Space-borne Networks}
\label{sec:network}

The space-borne gravitational wave detectors LISA \citep{Amaro-Seoane2017}, Taiji \citep{Hu2017}, and TianQin \citep{Luo2016} are all expected to be launched in the 2030s.  Given the overlap in their mission schedules, it is anticipated that networks could form between them \citep{Ruan2020a,Gong:2021gvw}, offering an excellent opportunity to detect parity-violating SGWB signals in the millihertz (mHz) frequency band \citep{Seto2020,Orlando2021}. 
In the following, we apply the formula derived in Section~\ref{sec:general} to the LISA-Taiji networks. We will examine their effective ORFs and sensitivity to the $I$ and $V$ components of an isotropic SGWB. Additionally, we discuss the potential optimal configuration of the network for detecting circularly polarized background signals.
\begin{figure}[ht]
  \centering
  \includegraphics[width=12cm]{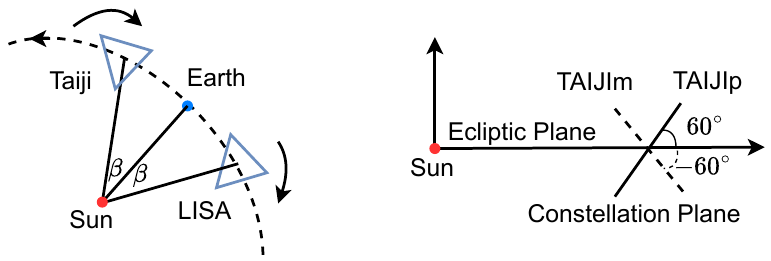} 
  \caption{Configuration of the LISA-Taiji network. LISA trails Earth with a separation angle of $\beta \sim 20^\circ$, while both TAIJIp and TAIJIm lead Earth by the same angle. The constellation planes of TAIJIp and TAIJIm are inclined at $+60^\circ$ and $-60^\circ$, respectively, with respect to the ecliptic plane.}
  \label{fig:orbit}
 \end{figure}

\subsection{(Alternative) LISA-Taiji networks}
\label{sec:lisataiji}
LISA and Taiji are both heliocentric detectors, with three spacecrafts forming a triangular constellation. The two detectors are planned to tail and lead Earth by an angle of $\beta \sim 20^\circ$, respectively. To ensure stable constellation formation, the planes of the constellations are designed to be inclined by about $\pm 60^\circ$ with respect to the ecliptic plane \citep{Dhurandhar2005}.
In addition to the fiducial case, labeled TAIJIp, \citep{Wang2021c} proposed two alternative orbits for Taiji: TAIJIm and TAIJIc. These alternative configurations are suggested to enhance the observation efficiency when combined with LISA in a network, improving sensitivity to both binary coalescence events and the isotropic SGWB~\citep{Wang2021c,Wang2021d}.

\begin{figure}[ht]
 \centering
 %\vspace {3.5cm}
 \includegraphics[width=7.5cm]{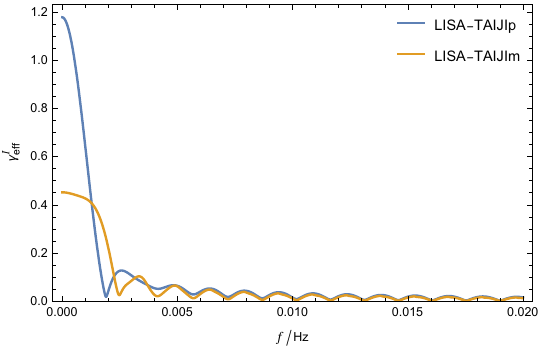}
 \includegraphics[width=7.5cm]{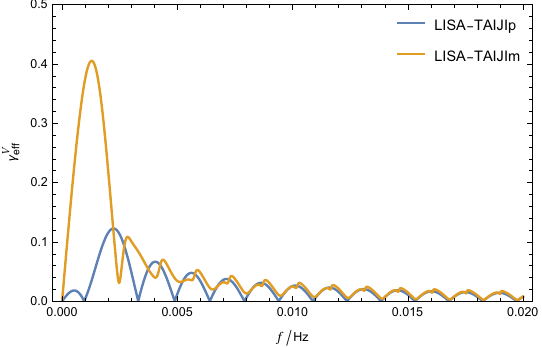}
 \caption{The effective overlap reduction functions of LISA-Taiji networks for  $I$ (left) and $V$ (right) components. The LISA-TAIJIp and LISA-TAIJIm configurations are indicated in blue and orange, respectively.}
 \label{fig:ORF}
\end{figure}

For detecting an isotropic parity-violating SGWB, the coplanar network LISA-TAIJIc is ineffective. Therefore, we focus on the TAIJIp and TAIJIm configurations. Both orbits lead Earth by an angle of $\beta \sim 20^\circ$, but differ in the inclination of their constellation planes relative to the ecliptic plane: $+60^\circ$ for TAIJIp and $-60^\circ$ for TAIJIm. An illustration of these orbits is provided in Fig.~\ref{fig:orbit}. With these orbit parameters, the angles between the separation vector and the normal vectors of each detector's constellation plane can be determined. For LISA-TAIJIp network, we have 
\begin{equation}
a = 1.27 , ~~~ b = 1.87 , ~~~ c = 0.60 .
\end{equation}
For the LISA-TAIJIm network, we have
\begin{equation}
a = 1.27 , ~~~ b = 1.87 , ~~~ c = 1.24 .
\end{equation}
Substitute these values into the formula given in Sec.~\ref{sec:ORF}, we obtain the effective ORFs for the two LISA-Taiji networks, as shown in Fig.~\ref{fig:ORF}. The blue line represents LISA-TAIJIp, and the orange line corresponds to LISA-TAIJIm.

For the effective ORF of the $I$ component, $\gamma_{\rm eff}^I(f)$, the LISA-TAIJIm configuration is only about one-third as sensitive as LISA-TAIJIp at the low-frequency limit, primarily due to the greater misalignment between the constellation planes of the two detectors. However, in the most sensitive frequency band around 2 mHz, the ORF of LISA-TAIJIm outperforms that of LISA-TAIJIp. At low frequencies, the ORF is dominated by the geometric contribution, along with the spherical Bessel function $j_0$, and decays rapidly as $\sim f^{-1}$ at $f \gg c/\Delta r$, where the phase coherence is lost due to the positional difference between the detectors. 
At higher frequencies, the ORFs of both configurations converge to $\sim f^{-1}$,
and the periodically wavy structure has a characteristic frequency of approximately half of $c/\Delta r \sim 3$ mHz, as determined by the spherical Bessel functions.
\begin{figure}[ht]
 \centering
 \includegraphics[width=7.5cm]{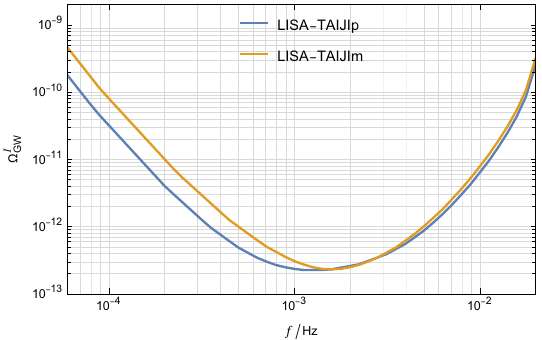}
 \includegraphics[width=7.5cm]{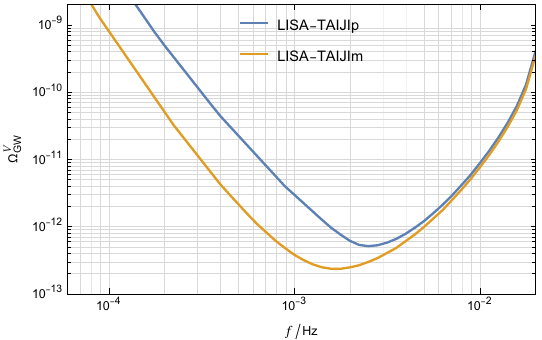}
 \caption{The power-law integrated sensitivity curve of LISA-Taiji networks for $I$ (left) and $V$ (right) components. The LISA-TAIJIp and LISA-TAIJIm configurations are indicated in blue and orange, respectively. The detection threshed is set to an SNR of 2, with an observation time of $T_{\rm obs}=3~\rm year$.}
 \label{fig:PLI}
\end{figure}

For the effect ORF of $V$ component $\gamma_{\rm eff}^V(f)$, the first thing to note is that both networks drop to zero in the low-frequency limit $f \rightarrow 0$~Hz. The ORF of LISA-TAIJIp remains quite small around $1~\rm mHz$, which explains why the SNR for the $V$ component is nearly unchanged as the low-frequency limit decreases below $2~\rm mHz$ in \citep{Seto2020a}. In contrast, for LISA-TAIJIm, the ORF peaks in the lower frequency regime around $f\sim 1~\rm mHz$ and is consistently larger than that of LISA-TAIJIp across almost all frequencies, except for a few small dips, primarily caused by the cross-term $\sum_\kappa \gamma^I_\kappa\gamma^V_\kappa$ in $\gamma_{\rm eff}^V$.

To illustrate the detectability of different network configurations, we also present the power-law integrated (PLI) sensitivity curve \citep{Thrane2013}, as a measure of the network's sensitivity to general power-law spectrum SGWBs. The PLI sensitivity curve provides a useful visualization of the detectable power-law SGWB signals across a wide range of possible spectral indices. It is particularly convenient for analyzing and comparing the capabilities of different detectors over broad frequency ranges. Following standard convention in \cite{Seto2006, Maggiore2007}, we use the fractional energy density spectrum $\Omega(f)$ in the sensitivity curve, which is related to the Stokes parameters $I(f), V(f)$ as follows:
\begin{equation}
  \label{eq:Omega}
  \Omega^{I}(f) = \frac{4\pi^2}{3H_0^2} f^3 I(f), ~~~~~ \Omega^{V}(f) = \frac{4\pi^2}{3H_0^2} f^3 V(f),
\end{equation}
where $H_0$ is the Hubble constant. For our calculations, we adopt $H_0=70 \rm km/s/Mpc$. 

With the effective ORFs, the PLI sensitivity can be readily obtained, as  described in appendix~\ref{sec:PIL}. The results are presented in Fig.~\ref{fig:PLI}, where the blue line represents LISA-TAIJIp, and the orange line corresponds to LISA-TAIJIm. Here, we adopt an observation time of $T_{\rm obs} = 3~\rm year$, and set the detection threshold to ${\rm SNR}_{\rm thr} = 2$, following \citep{Abbott2021b,Cornish2002}, corresponding to a 2$\sigma$ (95\% confidence level) detection \citep{Allen1996,Allen1999}.

For the overall intensity $I$ component, at lower frequencies, the sensitivity of LSA-TAIJIm is weaker compared to the default configuration, with the lower bound being approximately twice that of LISA-TAIJIp. However, the sensitivity becomes comparable to LISA-TAIJIp at the higher end of the frequency band, due to the superior ORF around 2mHz, as discussed earlier.
For the parity-violating $V$ component, the sensitivity of LSA-TAIJIm consistently outperforms that of LSA-TAIJIp, demonstrating an improvement of about an order of magnitude in the low-frequency regime. At higher frequencies, the sensitivity of both configurations are again nearly identical, consistent with the $\gamma_{\rm eff}^V(f)$ results discussed previously. Notably, with LISA-TAIJIm, the sensitivity of $V$ component is comparable to that of $I$ component in the most sensitive regime around 2mHz, providing an unparalleled opportunity to test various parity-violating theories in the mHz band.

\subsection{Discussion on the configurations}
\label{sec:discussion}

As discussed in section~\ref{sec:ORF}, the effective ORFs, as a function of frequency, can be decomposed into two components: the geometric part, $X_{mn}$, and the oscillatory part, $b_m(2 \pi f\frac{\Delta r}{c})$, represented in the form of Bessel functions. The geometric part is determined by the three angles between the constellation planes' normal vectors and the separation vector. The oscillatory part is primarily influenced by the separation $\Delta r$ between the two detectors.
To investigate the influence of different factors on the ORFs and identify the potential optimal configuration for detecting parity-violating SGWB, we will relax the constraints previously imposed on the LISA-Taiji networks. This allows us to explore more general network configurations and examine their implications for the detecting the parity-violating background.

\begin{figure}[ht]
  \centering
  \includegraphics[width=9.5cm]{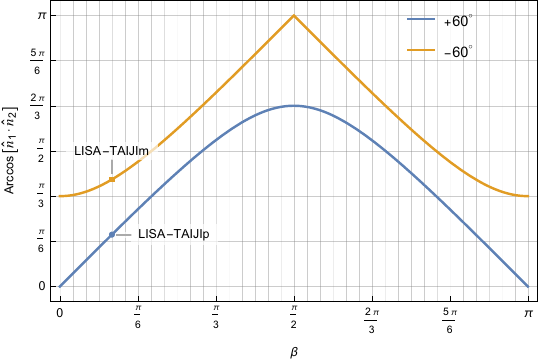}
\caption{The angle Arccos$[\hat{n}_1\cdot \hat{n}_2]$ between the normal vector of detectors changes with separation angle $\beta$ relative to the Earth. The blue curve shows the case both detectors' constellation planes keep $+60^\circ$ with the ecliptic plane. While, the orange curve shows the case of two detectors with $+60^\circ$ and $-60^\circ$ angles, separately. The previous networks LISA-TAIJIp and LISA-TAIJIm are shown as filled circle and square in the plot.}
  \label{fig:normal_vector_angle}
\end{figure}

Firstly, we maintain the inclination angle of the constellation plane with respect to the ecliptic plane while varying the separation angle $\beta$ of both detectors relative to Earth, deviating from the default $\beta \sim 20^\circ$. In this case, in addition to the change in separation distance, the relative orientation of the two detectors also changes. Fig.~\ref{fig:normal_vector_angle} shows the variation in the angle between the normal vector of the two detectors. Arccos$[\hat{n}_1\cdot \hat{n}_2]$ increase as $\beta$ grows, reaching its maximum at $\beta=\frac{\pi}{2}$. 
\begin{figure}[ht]
  \centering
  \includegraphics[width=9.5cm]{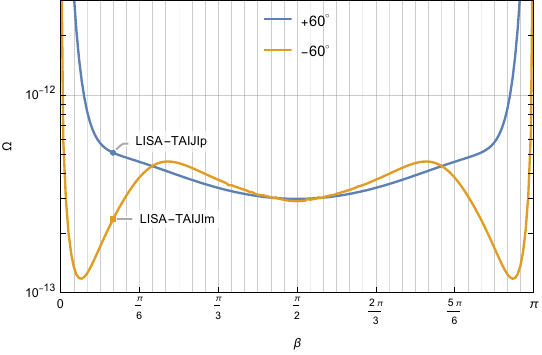}
  \caption{The sensitivity of detector networks to the $V$ component as a function of the separation angle $\beta$ relative to the Earth. Here, we assume a flat spectrum for the SGWB, with a detection threshold of ${\rm SNR}_{\rm thr} = 2$ and an observation time of $T_{\rm obs} = 3~\text{years}$. The blue curve represents the case where both detectors' constellation planes maintain an inclination of $+60^\circ$ with respect to the ecliptic plane, while the orange curve illustrates the scenario in which one detector is oriented at $+60^\circ$ and the other at $-60^\circ$. The previous networks, LISA-TAIJIp and LISA-TAIJIm, are depicted as filled circles and squares, respectively.}
  \label{fig:Omega_beta}
\end{figure}
\begin{figure}[ht]
  \centering
  \includegraphics[width=9.5cm]{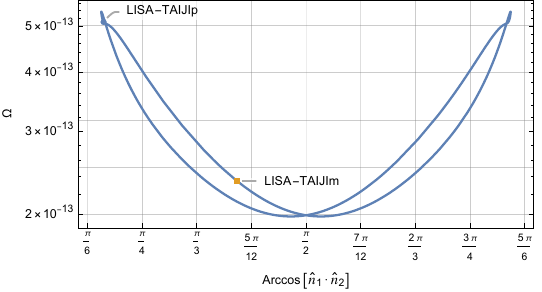}
  \caption{The sensitivity of detector networks to the $V$ component as a function of the angle of the normal vector of the detectors' constellation planes. Here, we assume a flat spectrum for the SGWB, with a detection threshold of ${\rm SNR}_{\rm thr} = 2$ and an observation time of $T_{\rm obs} = 3~\text{years}$. The previous networks, LISA-TAIJIp and LISA-TAIJIm, are represented as filled circles and squares, respectively.}
  \label{fig:Omega_normal_vector_angle}
\end{figure}

Assuming a flat spectrum of $\Omega(f)$, with a detection threshold of ${\rm SNR}_{\rm thr}=2$ and an observation time of $T_{\rm obs} = 3~\rm year$, the minimal detectable signal can be obtained from Eq.~\eqref{eq:SNR}. The lower bound of $\Omega$ as a function of angle $\beta$ is shown in Fig.~\ref{fig:Omega_beta}. The optimal configuration for $+60^\circ$ inclination of constellation plane occurs around $\beta\sim \frac{\pi}{2}$. 
While for $-60^\circ$ case, the optimal point is around $\beta\sim \frac{\pi}{25} \sim 7^\circ$, with a further gain of a factor of 2. 
% Around $\beta\sim \frac{\pi}{2}$, this corresponds to the two detectors being on opposite sides of the orbit with parallel constellation planes. Notably, this differs from the co-planar case, as the long-wavelength approximation—or more precisely, the small-antenna approximation—applies to a single detector. While, the distance between the detectors remains longer than the gravitational wavelength over most of the considered range.

From Fig.~\ref{fig:normal_vector_angle}, the constellation plane angle changes only slightly between the default value of $\beta\sim 20^\circ$ and $\beta\sim 7^\circ$, suggesting that the improvement in sensitivity for the $-60^\circ$ inclination case, compared to LISA-TAIJIm, primarily results from changes in the separation distance. At $\beta \sim 7^\circ$, the distance between the two detectors is about 2 to 3 times shorter than the default case of $\beta\sim 20^\circ$. This reduction shifts the peak frequency of the overlap reduction function (ORF) for the $V$ components to around 2 to 3 mHz, which coincides with the most sensitive frequency band of LISA and Taiji detectors.

Secondly, we fix the separation angle at $\beta=20^\circ$ and keep one detector's inclination angle with respect to the ecliptic plane constant, while varying the inclination angle of the other detector's constellation plane. In this scenario, the Bessel function part of the ORFs remains unchanged, and only the geometric part affects the overall sensitivity. The result is shown in Fig.~\ref{fig:Omega_normal_vector_angle}. From the figure, we observe that with a fixed separation angle, the sensitivity generally increases as the angle between the two detectors' constellation planes grows, with the most sensitive value occurring around $\frac{\pi}{2}$. This behavior can be understood by considering the parity-breaking nature of the $V$ component, which should be more prominent in configurations with greater symmetry breaking.

\section{Conclusion}
\label{sec:conclusion}

The circularly polarized gravitational wave backgrounds are predicted in many well-motivated models of inflation and phase transition involving spontaneous parity violation. The (non-)detection of such parity-violating signals would be a new approach to probing the nature of gravity and imposing constraints on various parity-violating theories.
In this paper, we study the detection of a parity-violating isotropic SGWB with a network of two space-borne triangular detectors. We first derive the general analytical formula for the ORFs of networks composed of any two triangular detectors, utilizing the symmetry of the system under the long-wave approximation.

Based on the  analytical results, we evaluate the detectability of a parity-violating SGWB with alternative LISA-Taiji network configurations. We find that by changing the orientation of Taiji's constellation plane relative to the ecliptic plane from $+60^\circ$ to $-60^\circ$, the sensitivity to the $V$ component at low frequencies is significantly enhanced, while the sensitivity to the $I$ component is only slightly reduced. Notably, the sensitivity gain for the $V$ component is approximately one order of magnitude in the frequency region around millihertz, making its peak sensitive comparable to that of the $I$ component. This provides us a great opportunity to test various parity-violating theories using the LISA-Taiji network.

% The result can be straightforwardly applied to networks with other triangle detectors like TianQin and Einstein Telescope.

We also provide a general discussion on the optimal configuration by relaxing the constraints imposed on the LISA-Taiji networks. Our findings suggest that by maintaining the $-60^\circ$ inclination of Taiji's constellation plane relative to the ecliptic plane, as in the TAIJIm configuration, the peak sensitivity can be further improved by a factor of 2 with a smaller angle of $\beta \sim 7^\circ$. While keeping the separation angle fixed, the sensitivity of the $V$ component generally increases as the symmetry breaking of the network grows, with the peak sensitivity occurring when the constellation planes are nearly orthogonal. These insights could be valuable for the design of future detector networks.

One thing to note is that the sensitivity depends on the limit of the frequency integration in equation \eqref{eq:SNR}. In this work, we adopt the long-wave approximation and set the upper limit to $f_{\rm max} \sim c/(2\pi L) \sim 20~\rm mHz$, which should be a valid choice. The lower frequency limit, on the other hand, is mainly constrained by the subtraction of the confusion foreground produced by galactic and extragalactic binary systems, which, in turn, rely on the observation time $T_{\rm obs}$, as discussed in previous work \citep{Robson2019}.

Finally, it is important to note that the results of our work relay on several assumptions. We employ the long-wave approximation, which should be valid as long as we focus on the low-frequency regime. Additionally, We assume that the $T$ channel can be neglected. This assumption may not hold in practice due to the influence of unequal and unstable arm lengths. 
More detailed numerical calculations with realistic orbits and robust TDI scheme have been done in \citep{Chen:2024fto}.

%More detailed numerical calculations are currently in preparation.

% \vspace{+50pt}

\begin{acknowledgments}
We thank many valuable discussions with Gang Wang. This work is supported by the National Key Research and Development Program of China Grant (No.2021YFC2201901, No.2023YFC2206200), the National Natural Science Foundation of China Grants (No.12375059, No.12147132, No.12405074), and the Fundamental Research Funds for the Central Universities(No.E2ET0209X2). 
\end{acknowledgments}

\appendix

\section{AET Channels of Space-borne Detectors}
  \label{sec:AET}

\begin{figure}[ht]
  \centering
  %\vspace {3.5cm}
  \includegraphics[width=8.5cm]{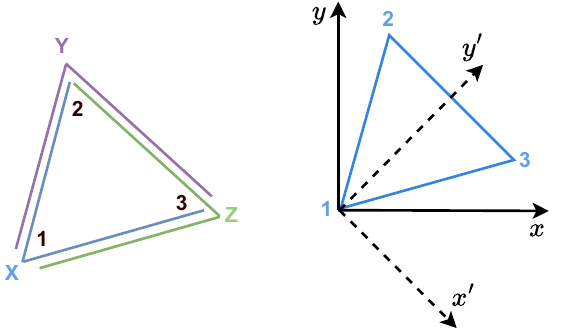}
  \caption{The left panel illustrates the geometry of a LISA-like space interferometer and the associated laser arm links used for generating each of the Michelson-type TDI channels $X$, $Y$, and $Z$. The right panel presents the coordinate system associated with the noise-orthogonal TDI channels $A$ and $E$, relative to the spacecraft constellation.}
   \label{fig:AET}
\end{figure}

The time-delay interferometry (TDI) is essential for space-borne interferometers to achieve the targeted sensitivity. To effectively suppress the laser noise, various TDI channels are constructed by combining measurements from different inter-spacecraft laser arm links. The noise-orthogonal signals, $A$ and $E$, used in this work are constructed from the Michelson-type TDI generators $X, Y, Z$, as shown in the left panel of Fig.~\ref{fig:AET}.
Under the assumption of equal arm lengths and the long-wavelength approximation, the Michelson-type signal in terms of fractional frequency deviations extracted from vertex 1 is given by \citep{Babak2023}:
\begin{equation}
  \tilde{X}_{1.5} = -4 \frac{f}{f_*} \sin \frac{f}{f_*} ~ \frac{\hat{r}^a_{12}\hat{r}^b_{12} - \hat{r}^a_{13}\hat{r}^b_{13}}{2}\tilde{h}_{ab}(f),
\end{equation}
where $\hat{r}_{ij}$ is the unit vector of laser link connects spacecraft $i$ and $j$, $f_* \equiv c/(2\pi L)$ and $L$ is the arm-length. The expressions for $Y$ and $Z$ follow by cyclic permutation of the spacecraft labels \{1, 2, 3\} in the above expression of $X$.

The noise power spectral density (PSD) of the $X$, $Y$ and $Z$ channels are the same assuming that the PSD of the different links are all equal, which is given by \citep{Vallisneri2012,Babak2021}:
\begin{equation}
  N_X(f)  = \frac{16}{L^2} \left(\frac{f}{f_*}  \sin \frac{f}{f_*} \right)^2 \left[P_{\rm OMS} + 2\left(1+\cos^2 \frac{f}{f_*}\right)\frac{P_{\rm acc}}{(2\pi f)^4} \right].
\end{equation}
Also noise is correlated between $X$, $Y$ and $Z$, with the cross spectral density \citep{Vallisneri2012}:
\begin{equation}
  N_{XY}(f) = N_{YZ}(f) = N_{XZ}(f)  = -\frac{8}{L^2} \left(\frac{f}{f_*}  \sin \frac{f}{f_*} \right)^2  \cos \frac{f}{f_*} \left[P_{\rm OMS} + 4 \frac{P_{\rm acc}}{(2\pi f)^4} \right].
\end{equation}
The optical measurement noise $P_{\rm OMS}$ and acceleration noise $P_{\rm acc}$ are given by:
\begin{eqnarray}
    P_{\rm OMS} &=& A^2_{\rm OMS}  \left[1 + \left(\frac{2 \rm mHz}{f}\right)^4\right],\\
    P_{\rm acc} &=& A^2_{\rm acc} \left[1 + \left(\frac{0.4 \rm mHz}{f}\right)^2\right]\left[1 + \left(\frac{f}{8 \rm mHz}\right)^2\right],
\end{eqnarray}
where $A_{\rm OMS}$ is the noise budget of the optical measurement system, and $A_{\rm acc}$ is the noise budget of acceleration of test mass. For LISA, the noise budgets are $A_{\rm OMS} = 15 {\rm pm}/\sqrt{\rm Hz}$ and $A_{\rm acc} = 3 {\rm fm/s^2}/\sqrt{\rm Hz}$ \citep{Babak2021,Colpi2024}. For Taiji, the noise budgets are $A_{\rm OMS} = 8 {\rm pm}/\sqrt{\rm Hz}$ and $A_{\rm acc} = 3 {\rm fm/s^2}/\sqrt{\rm Hz}$ \citep{Luo2020,Luo2021a}.

Following \citep{Adams2010}, we use the following combinations to obtain noise-orthogonal TDI channels $A$, $E$, and $T$:
\begin{equation}
    A = \frac{2X-Y-Z}{3}, ~~ E = \frac{Z-Y}{\sqrt{3}}, ~~ T = \frac{X+Y+Z}{3}.
\end{equation}
Given that the $T$ channel has significantly lower sensitivity in the low-frequency regime \citep{Prince2002}, we omit it in this work. Converting from relative frequency shift to the strain response leads to the following expressions for the $A$ and $E$ channels:
\begin{equation}
  A = \frac{\sqrt{3}}{2}  \sin \frac{f}{f_*}~ D^{ab}_A~ \tilde{h}_{ab}(f), ~~~~ 
E = \frac{\sqrt{3}}{2} \sin \frac{f}{f_*}~ D^{ab}_E~ \tilde{h}_{ab}(f),
\end{equation}
where $D^{ab}_A$ and $D^{ab}_E$ are the detector tensors for channels $A$ and $E$, respectively, defined as:
\begin{equation}
  D^{ab}_A = \frac{\hat{x}^a\hat{x}^b - \hat{y}^a\hat{y}^b}{2},~~~~ D^{ab}_E = \frac{\hat{x}'^a\hat{x}'^b - \hat{y}'^a\hat{y}'^b}{2}.
\end{equation}
Here $\hat{x}$, $\hat{y}$, and $\hat{x}'$, $\hat{y}'$ are the unit vectors defined in the coordinates shown in the right panel of Fig.~\ref{fig:AET}. These detector tensors indicate that, under the long-wave approximation, the $A$ and $E$ channels effectively behave as two L-shaped Michelson interferometers that are misaligned by $45^\circ$ in the constellation plane \citep{Seto2020a}. Notably, due to the orbital motion of the spacecraft, these channels rotate within the constellation plane over the course of a year, a movement known as ``cartwheel motion".

It also follows that the noise PSD of A and E channels, which is identical, is given by:
\begin{equation}
  \begin{aligned}
    N(f)  = \frac{2}{3}\frac{\sin^2 \frac{f}{f_*}}{L^2}\Bigg[&\left(1+\frac{1}{2} \cos \frac{f}{f_*}\right)P_{\rm OMS} %\\&
    + 2\left(1+\cos \frac{f}{f_*} +\cos^2 \frac{f}{f_*}\right)\frac{P_{\rm acc}}{(2\pi f)^4} \Bigg].
  \end{aligned}
\end{equation}
% Compare to use in \citep{Seto2020a}, a factor of $\frac{\sqrt{3}}{2}\sin x$: $\frac{\sqrt{3}}{2}$ is due to the $60^\circ$ angle of the arms, while $\sin \frac{f}{f_*}$ come from the TDI operate.

\section{The Analytical Formulas for Overlap Reduction Functions}
\label{sec:ORFtensor}
With the normalization introduced in Eq.~\eqref{eq:orf}, we have:
\begin{equation}
  \begin{aligned}
    \gamma^I_{ij}(f) &= \frac{5}{2} D_i^{~ab} D_j^{~cd} \int \frac{d^2\hat{k}}{4\pi} ~ \left[e^+_{ab}(\hat{k}) e^+_{cd}(\hat{k}) + e^\times_{ab}(\hat{k}) e^\times_{cd}(\hat{k})\right]  e^{-\ii 2\pi f\hat{k}\cdot \Delta\vec{r}/c},\\
    \gamma^V_{ij}(f) &=  -\ii ~\frac{5}{2} D_i^{~ab} D_j^{~cd} \int \frac{d^2\hat{k}}{4\pi} \left[e^+_{ab}(\hat{k}) e^\times_{cd}(\hat{k}) - e^\times_{ab}(\hat{k}) e^+_{cd}(\hat{k})\right]  e^{-\ii 2\pi f\hat{k}\cdot \Delta\vec{r}/c}.
  \end{aligned}
\end{equation}
The integrals can be done analytic as shown in \citep{Flanagan1993,Allen1999,Seto2008}. Here we give a brief introduction. For simplification of the notation in the following, we introduce the unit vector of separation of two detectors $\hat{s}$, and define:
\begin{equation}
  \hat{s} \equiv -\Delta \vec{r}/\Delta r, ~~~~ \iota \equiv \frac{2 \pi f \Delta r}{c}.
\end{equation}
Following \citep{Flanagan1993,Allen1999}, we introduce the auxiliary tensors:
\begin{equation}
  \label{eq:gamma_tensor}
  \begin{aligned}
    \Gamma^I_{abcd}(\iota, \hat{s}) &\equiv \frac{5}{2} \int \frac{d^2\hat{k}}{4\pi} ~ \left[e^+_{ab}(\hat{k}) e^+_{cd}(\hat{k}) + e^\times_{ab}(\hat{k}) e^\times_{cd}(\hat{k})\right]  e^{\ii \iota \hat{k}\cdot \hat{s}},\\
    \Gamma^V_{abcd}(\iota, \hat{s}) &\equiv  -\ii ~\frac{5}{2} \int \frac{d^2\hat{k}}{4\pi} \left[e^+_{ab}(\hat{k}) e^\times_{cd}(\hat{k}) - e^\times_{ab}(\hat{k}) e^+_{cd}(\hat{k})\right]  e^{\ii \iota \hat{k}\cdot \hat{s}}. 
  \end{aligned}
\end{equation}
% unit vector that points in the direction connecting the two detectors
One can utilize the symmetric property of $\Gamma^I_{abcd}$ and $\Gamma^V_{abcd}$ to get their analytic formulas, without directly calculating the integration, as shown in the following. The normalized ORFs are then followed by:
\begin{equation}
  \begin{aligned}
  \label{eq:gamma_decomposed}
  \gamma^I_{ij}(f) & = D_i^{~ab} D_j^{~cd}\Gamma^I_{abcd}(\iota, \hat{s}) , \\ \gamma^V_{ij}(f) &=  D_i^{~ab} D_j^{~cd}\Gamma^V_{abcd}(\iota, \hat{s}).
    \end{aligned}
\end{equation}

\subsection{Parity even case}
From the definition in Eq.~\eqref{eq:gamma_tensor}, one can find that $\Gamma^I_{abcd}(\iota,\hat{s})$ is a symmetric tensor under the interchanges of indices $a\leftrightarrow b$, $c\leftrightarrow d$, and $ab\leftrightarrow cd$. It is also trace-free concerning the $ab$ and $cd$ index pairs. With the constraint of all these properties, the most general form of the tensor can be constructed with $\delta_{ab}$ and $s_a$:
\begin{equation}
  \begin{aligned}
    \Gamma^I_{abcd}(\iota, \hat{s}) =& ~ A(\iota)\delta_{ab}\delta_{cd} +  B(\iota)(\delta_{ac}\delta_{bd} + \delta_{bc}\delta_{ad}) + C(\iota)(\delta_{ab}s_c s_d + \delta_{cd}s_a s_b)\\
    & + D(\iota)(\delta_{ac}s_b s_d + \delta_{ad}s_b s_c + \delta_{bc}s_a s_d + \delta_{bd}s_a s_c) + E(\iota) s_a s_b s_c s_d.
  \end{aligned}
  \label{eq:tensor_decomposed}
\end{equation}
$\Gamma^I_{abcd}(\iota, \hat{s})$ is a combination of five base tensors $(\delta^{ac}\delta^{bd}+\delta^{bc}\delta^{ad})$, $\delta^{ab}\delta^{cd}$, $\cdots$, $s^a s^b s^c s^d$ with the coefficients $A(\iota)$, $B(\iota)$, $C(\iota)$, $D(\iota)$, and $E(\iota)$ as some undetermined function of $\iota$. Contract this tensor with each of the bases, one can get the following linear system of equations for these coefficients:
\begin{equation}
  \begin{pmatrix}
    9  &  6  & 6 & 4  & 1\\
    6  & 24  & 4 & 16 & 1\\
    6  &  4  & 8 & 8  & 4\\
    4  & 16  & 8 & 24    & 2\\
    1  &  2  & 2 & 4    & 1
  \end{pmatrix}
  \begin{pmatrix}
    A\\
    B\\
    C\\
    D\\
    E
  \end{pmatrix}(\iota) =
  \begin{pmatrix}
    p^I_0\\
    p^I_1\\
    p^I_2\\
    p^I_3\\
    p^I_4
  \end{pmatrix}(\iota),
\label{eq:ABCDE}
\end{equation}
where
\begin{equation}
  \begin{aligned}
    p^I_0(\iota) &\equiv \Gamma^I_{abcd}(\iota,\hat{s})\delta^{ab}\delta^{cd},\\
    p^I_1(\iota) &\equiv \Gamma^I_{abcd}(\iota,\hat{s})(\delta^{ac}\delta^{bd}+\delta^{bc}\delta^{ad}),\\
    p^I_2(\iota) &\equiv \Gamma^I_{abcd}(\iota,\hat{s})(\delta^{ab}s^c s^d + \delta^{cd}s^a s^b),\\
    p^I_3(\iota) &\equiv \Gamma^I_{abcd}(\iota,\hat{s})(\delta^{ac}s^b s^d+\delta^{ad}s^b s^c+\delta^{bc}s^a s^d+\delta^{bd}s^a s^c),\\
    p^I_4(\iota) &\equiv \Gamma^I_{abcd}(\iota,\hat{s})s^a s^b s^c s^d.
  \end{aligned}
\end{equation}

Substitute the definition of $\Gamma^I_{abcd}(\iota,\hat{s})$ given in Eq.~\eqref{eq:gamma_tensor}, one can get the $p^I$ in integral form. As a scalar, these integrals can be easily evaluated without loss of generality after choosing some special coordinate system. Here we choose the coordinate system such that the unit vector $\hat{s}$ along the $z$ axis. Following \citep{Allen1999}, we set the polarization tensors to:
\begin{equation}
  e^+_{ab}(\hat{k})=\hat{m}_a \hat{m}_b-\hat{n}_a \hat{n}_b, \qquad
  e^\times_{ab}(\hat{k})=\hat{m}_a \hat{n}_b + \hat{n}_a \hat{m}_b,
\end{equation}
with orthogonal set of unit vectors $\hat{m}, \hat{n}, \hat{k}$ given by:
\begin{equation}
\begin{aligned}
  \hat{k} &= \cos\phi\sin\theta~ \hat{x} +  \sin\phi\sin\theta~ \hat{y} + \cos\theta~ \hat{z},\\
  \hat{m} &= \sin\phi~ \hat{x} -  \cos\phi~ \hat{y},\\
  \hat{n} &= \cos\phi\cos\theta~ \hat{x} +  \sin\phi\cos\theta~ \hat{y} - \sin\theta~ \hat{z},\\
\end{aligned}
\end{equation}
where $(\theta, \phi)$ are the standard polar and azimuth angles of spherical coordinate. Then 
\begin{equation}
\hat{k}\cdot\hat{s}=\cos\theta\ ,\quad
\hat{m}\cdot\hat{s}=0\ ,\quad
\hat{n}\cdot\hat{s}=-\sin\theta\ , \label{theta}
\end{equation}
and the integrals can be simplified as:
\begin{equation}
  \begin{aligned}
    p^I_0(\iota) &= 0,\\
    p^I_1(\iota) &= 10\int_{-1}^1 dx\ e^{i\iota x}=20 ~j_0(\iota),\\
    p^I_2(\iota) &= 0,\\
    p^I_3(\iota) &= 10\int_{-1}^1 dx\ e^{i\iota x}\ (1-x^2) = \frac{40}{\iota} ~j_1(\iota),\\
    p^I_4(\iota) &= \frac{5}{4}\int_{-1}^1 dx\ e^{i\iota x}\ (1-x^2)^2 = \frac{20}{\iota^2} ~j_2(\iota),
  \end{aligned}
\end{equation}
where $j_0(\iota)$, $j_1(\iota)$, and $j_2(\iota)$ are spherical Bessel functions of the first kind:
\begin{equation}
  \begin{aligned}
    j_0(\iota) &= \frac{\sin\iota}{\iota},\\
    j_1(\iota) &= \frac{\sin\iota}{\iota^2} - \frac{\cos\iota}{\iota},\\
    j_2(\iota) &= 3~\frac{\sin\iota}{\iota^3} - 3~\frac{\cos\iota}{\iota^2} -\frac{\sin\iota}{\iota}.
  \end{aligned}
\end{equation}
The higher-order spherical Bessel function follows as 
\begin{equation}
j_{n+1}(\iota) = \frac{2n+1}{\iota} j_{n}(\iota) - j_{n-1}(\iota).
\end{equation}

The coefficients $A(\iota)$, $B(\iota)$, $C(\iota)$, $D(\iota)$, and $E(\iota)$ as function of $\iota$ can then be solved from Eq.~\eqref{eq:ABCDE}, and we have:
\begin{equation} \label{eq:Icoefficients}
  \begin{pmatrix}
    A\\
    B\\
    C\\
    D\\
    E
  \end{pmatrix}(\iota) = 
  \frac{5}{2}
  \begin{pmatrix}
    -1  & ~~2  &  ~~1\\
    ~~1 &  -2  &  ~~1\\
    ~~1 &  -2  &   -5\\
    -1  & ~~4  &   -5\\
    ~~1 & -10  & ~~35\\
  \end{pmatrix}
  \begin{pmatrix}
    j_0\\
    \iota^{-1} j_1\\
    \iota^{-2} j_2\\
  \end{pmatrix}(\iota).
\end{equation}
Substitute \eqref{eq:Icoefficients} into (\ref{eq:gamma_decomposed}), we can get the analytic form of $\gamma_{ij}^I(f)$. Noticing that the detector tensors $D_i^{ab}$ and $D_j^{cd}$ are both trace free, we have:
\begin{equation}
  \gamma^I_{ij}(f)= D^{ab}_i D^{cd}_j \big[2A(\iota)~ \delta_{ac}\delta_{bd} + 4C(\iota) ~\delta_{ac} s_b s_d + E(\iota)~ s_a s_b s_c s_d\big].
\end{equation}
Here we introduce new auxiliary tensor $\hat{\Gamma}^I_{abcd}(\iota, \hat{s})$ as a simplified form of $\Gamma^I_{abcd}(\iota, \hat{s})$, and retain the following relation:
\begin{equation}
  \gamma^I_{ij}(f) = D^{ab}_i D^{cd}_j \hat{\Gamma}^I_{abcd},
\end{equation}
with
\begin{equation}
  \hat{\Gamma}^I_{abcd} \equiv b^I_0(\iota)\delta_{ac}\delta_{bd} + b^I_1(\iota) ~\delta_{ac} s_b s_d + b^I_2(\iota)~ s_a s_b s_c s_d,
\end{equation}
where 
\begin{equation}
  \begin{pmatrix}
    b_0\\
    b_1\\
    b_2
  \end{pmatrix} \equiv 
  \begin{pmatrix}
    2B\\
    4D\\
    E
  \end{pmatrix} = 
  \frac{5}{2}
  \begin{pmatrix}
    ~~2 & -4   & ~2\\
    -4  & ~~16 & -20\\
    ~~1 & -10  & ~35
  \end{pmatrix}
  \begin{pmatrix}
    j_0\\
    \iota^{-1} j_1\\
    \iota^{-2} j_2\\
  \end{pmatrix} = 
  \frac{1}{7}
  \begin{pmatrix}
    14  & -21  & ~~1\\
    ~0  & ~60  & -10\\
    ~0  & ~~0  & ~35/2
  \end{pmatrix}
  \begin{pmatrix}
    j_0\\
    j_2\\
    j_4\\
  \end{pmatrix}.
\end{equation}
Here $j_4(\iota)$ is the spherical Bessel function of order 4.

\subsection{Parity odd case}
Similarly, for $\Gamma^V_{abcd}(\iota,\hat{s})$ given in Eq.~\eqref{eq:gamma_tensor}, one can find that it is a symmetric tensor under the interchanges of indices $a\leftrightarrow b$, $c\leftrightarrow d$, while antisymmetric tensor with $ab\leftrightarrow cd$. It is also trace-free with respect to the $ab$ and $cd$ index pairs. With antisymmetric tensor $w_{bd}\equiv\epsilon_{abc}s^c$, the general form of tensor following the constraints can be constructed as:
\begin{equation}
  \begin{aligned}
    \Gamma_{abcd}^V(\iota,\hat{s}) =~ & F(\iota)~ (\delta_{ab}s_c s_d - \delta_{cd}s_a s_b)\\
    & + G(\iota)~ (w_{ac} \delta_{bd}+ w_{ad}\delta_{bc}+w_{bc}\delta_{ad}+ w_{bd}\delta_{ac})\\
    & + H(\iota)\ (w_{ac}s_b s_d + w_{ad}s_b s_c + w_{bc}s_a s_d + w_{bd}s_a s_c).
  \end{aligned}
\end{equation}
Contract this tensor with each of the basis, one can get the following linear system of equations:
\begin{equation}
  \begin{pmatrix}
    0  & 4 & 0\\
    40 & 0 & 8\\ 
    8  & 0 & 8
  \end{pmatrix}
  \begin{pmatrix}
    F\\
    G\\
    H
  \end{pmatrix}(\iota) =
  \begin{pmatrix}
    p^V_0\\
    p^V_1\\
    p^V_2\\
  \end{pmatrix}(\iota),
\label{eq:FGH}
\end{equation}
where
\begin{equation}
  \begin{aligned}
    p^V_0(\iota)&\equiv \Gamma^V_{abcd}(\iota,\hat{s})(\delta^{ab}s^c s^d - \delta^{cd}s^a s^b),\\
    p^V_1(\iota)&\equiv \Gamma^V_{abcd}(\iota,\hat{s})(w^{ac} \delta^{bd} + w^{ad}\delta^{bc} \
                         + w^{bc}\delta^{ad} + w^{bd}\delta^{ac}),\\
    p^V_2(\iota)&\equiv \Gamma^V_{abcd}(\iota,\hat{s})(w^{ac}s^b s^d+w^{ad}s^b s^c +w^{bc}s^a s^d+w^{bd}s^a s^c).
  \end{aligned}
\end{equation}
With the same coordinate system for parity even case, we have
\begin{equation}
  \begin{aligned}
    p^V_0(\iota)& = 0,\\
    p^V_1(\iota)& = -\ii~ 20 \int_{-1}^1 dx\ e^{i\iota x} ~ x= 40~ j_1(\iota),\\
    p^V_2(\iota)& = -\ii~ 10\int_{-1}^1 dx\ e^{i\iota x} ~ x(1-x^2) = \frac{40}{\iota}~ j_2(\iota),\\
  \end{aligned}
\end{equation}
The coefficients $F(\iota)$ , $G(\iota)$, and $H(\iota)$ are following as solving Eq.~\eqref{eq:FGH}, with $F(\iota) =0$ and
\begin{equation}
  \begin{pmatrix}
    G\\
    H
  \end{pmatrix} =
  \frac{5}{4}
  \begin{pmatrix}
    ~~1 & -1\\
    -1  & ~~5
  \end{pmatrix}
  \begin{pmatrix}
    j_1\\
    \iota^{-1} j_2\\
  \end{pmatrix}.
\end{equation}

Again with the trace free property of detector tensors $D_i^{ab}$ and $D_j^{cd}$ are both trace free, we have:

\begin{equation}
  \gamma^V_{ij}(f)= D^{ab}_i D^{cd}_j \big[4G(\iota)~ w_{ac}\delta_{bd} + 4H(\iota) ~w_{ac} s_b s_d\big].
\end{equation}
And we introduce $\hat{\Gamma}^V_{abcd}(\iota, \hat{s})$ as the simplified form of $\Gamma^V_{abcd}(\iota, \hat{s})$:
\begin{equation}
  \gamma^V_{ij}(f) = D^{ab}_i D^{cd}_j \hat{\Gamma}^V_{abcd},
\end{equation}
with
\begin{equation}
  \hat{\Gamma}^V_{abcd} \equiv b^V_0(\iota)~ w_{ac}\delta_{bd} + b^V_1(\iota) ~w_{ac} s_b s_d,
\end{equation}
where 
\begin{equation}
  \begin{pmatrix}
    b^V_0\\
    b^V_1\\
  \end{pmatrix} \equiv 
  \begin{pmatrix}
    4G\\
    4H
  \end{pmatrix} =
  5\begin{pmatrix}
    ~~1 & -1\\
    -1  & ~~5
  \end{pmatrix}
  \begin{pmatrix}
    j_1\\
    \iota^{-1} j_2\\
  \end{pmatrix} = 
  \begin{pmatrix}
    4 & -1\\
    0  & ~~5
  \end{pmatrix}
  \begin{pmatrix}
    j_1\\
    j_3\\
  \end{pmatrix}.
\end{equation}
Here $j_3(\iota)$ is the spherical Bessel function of order 3.

\section{Power-Law Integrated Sensitivity Curve}
\label{sec:PIL}
The power-law integrated (PLI) sensitivity curve, first introduced in \citep{Thrane2013}, is one of the most commonly used methods to represent the sensitivity of detectors to a SGWB with a general power-law power spectrum. It can be obtained using the SNR provided in Eq.~\eqref{eq:SNR}, along with the effective ORFs discussed earlier.

SGWB signals characterized by a power-law spectrum can be described as $\Omega(f) = \Omega_\alpha\left(\frac{f}{f_c}\right)^\alpha$, where $\alpha$ is the spectral index, and $\Omega_\alpha$ is the value of the spectrum at a reference frequency $f_c$. The PSD of the background is then get from Eq.~\eqref{eq:Omega} and substituted into Eq.~\eqref{eq:SNR}, allowing us to determine the minimal detectable value of $\Omega_\alpha$, given a specified detection threshold for the SNR:
\begin{equation}
  \Omega^{\{I,V\}}_{\alpha_{\rm thr}} =  \frac{10}{3} \frac{{\rm SNR}_{\rm thr}}{\sqrt{T_{\rm obs}}} \frac{4\pi^2}{3H_0^2} \left[\int df \frac{\gamma_{\rm eff}^{\{I,V\}^2}(f) f^{2\alpha-6}}{N^2(f)}\right]^{-\frac{1}{2}} f_c^\alpha\,.
  \label{eq:Omega_limit}
\end{equation}
For each specific spectral index $\alpha$, the lower bound $\Omega_{\alpha_{\rm thr}}$ and the corresponding fractional energy density spectrum can be determined. The PLI sensitivity is then defined as the upper envelope of these spectra across all possible spectral indices.
\begin{equation}
  \Omega^{\{I,V\}}_{\rm PLI}(f) = \max_{\alpha}\left[\Omega^{\{I,V\}}_{\alpha_{\rm thr}} \left(\frac{f}{f_c}\right)^\alpha\right].
\end{equation}

\bibliographystyle{JHEP}
\bibliography{stochastic}
%%%%%%%%%%%%%%%%%%%%%%%%%%%%%%%%%%%%%%%%%%%%%%%%%
%\newpage

\end{document}